\documentstyle[aaspptwo,epsf]{article}   

\newcommand{\exvalt}[1]{\mbox{$<$}\mbox{#1}\mbox{$>$}}  
\newcommand{\um}{~$\mu$m}

\received{2 December 1994}
\accepted{25 May 1995}
\journalid{453}{10 November 1995}
\articleid{001}{999}
\slugcomment{To appear in the Astrophysical Journal, 10 November 1995}

\begin{document}




\title{Multiwavelength Energy Distributions and Bolometric Luminosities of the
12--Micron Galaxy Sample\footnote{This paper has been accepted for publication
in the 1995 November 10 issue of ApJ.}$^,$\footnote{The three data tables
mentioned in this paper (which do not appear in this preprint) can be accessed
via anonymous ftp at eggneb.astro.ucla.edu in the /pub/rush directory. These
ASCII files are named Spinoglio95-Table\{1,2,3\}.dat. These data tables will
also be included in the AAS CD--ROM Series ``Astrophysics on Disc", Volume~V,
1995.}}

\author{Luigi Spinoglio} \affil{Istituto di Fisica dello Spazio
Interplanetario,
CNR\\ CP 27 00044 Frascati, Italy\\ Electronic Mail:
luigi@orion.ifsi.fra.cnr.it}

\author{Matthew A. Malkan, Brian Rush} \affil{Department of Physics and
Astronomy, University of California at Los Angeles\\ Los Angeles, CA
90024--1562\\ Electronic Mail: malkan@bonnie.astro.ucla.edu;
rush@bonnie.astro.ucla.edu}


\author{Luis Carrasco\altaffilmark{3}, Elsa Recillas--Cruz\altaffilmark{3}}
\affil{Instituto Nacional de Astrof\'{\i}sica, Optica y Electr\'onica,\\
Apartado Postal 51, Puebla, Pue., M\'exico, CP 72000\\ Electronic Mail:
carrasco@tonali.inaoep.mx; elsare@tonali.inaoep.mx}

\altaffiltext{3}{On sabbatical leave from UNAM, M\'exico}

\addtocounter{footnote}{1} 

\begin{abstract}

Aperture photometry from our own observations and the literature is presented
for the 12\um\ Galaxies in the near infrared J, H and K bands and, in some
cases, in the L band. These data are corrected to ``total'' near--infrared
magnitudes, (with a typical uncertainty of 0.3 magnitudes) for a direct
comparison with our IRAS fluxes which apply to the entire galaxy. The corrected
data are used to derive integrated total NIR and FIR luminosities. We then
combine these with blue photometry and an estimate of the flux contribution
from
cold dust at wavelengths longward of 100\um\ to derive the first {\it
bolometric\/} luminosities for a large sample of galaxies.

The presence of nonstellar radiation at 2---3\um\ correlates very well with
nonstellar IRAS colors. This enables us to identify a universal Seyfert~nuclear
continuum from near-- to far--infrared wavelengths. Thus there is a sequence of
infrared colors which runs from a pure ``normal galaxy" to a pure
Seyfert/quasar
nucleus. Seyfert~2 galaxies fall close to this same sequence, although only a
few extreme narrow--line Seyferts have quasar--like colors, and these show
strong evidence of harboring an obscured Broad Line Region. A corollary is that
the host galaxies of Seyfert~nuclei have normal near-- to far--infrared spectra
on average. Starburst galaxies lie significantly off the sequence, having a
relative excess of 60\um\ emission probably due to stochastically--heated dust
grains. We use these correlations to identify several combinations of infrared
colors which discriminate between Seyfert~1 and~2 galaxies, LINERs, and
ultraluminous starbursts. In the infrared, Seyfert~2 galaxies are much more
like
Seyfert~1's than they are like starbursts, presumably because both kinds of
Seyferts are heated by a single central source, rather than a distributed
region
of star formation.

Moreover, combining the [25---2.2\um] color the [60---12\um] color, it appears
that Seyfert~1 galaxies are segregated from Seyfert~2s and starburst galaxies
in
a well-defined region characterized by the hottest colors, corresponding to the
flattest spectral slopes. Virtually no Seyfert~2 galaxy is present in such a
region. To reconcile this with the ``Unified Scheme" for Seyfert~1s and~2s
would
therefore require that the higher frequency radiation from the nuclei of
Seyfert~2s to be absorbed by intervening dust and re--emitted at lower
frequencies.

We find that bolometric luminosity is most closely proportional to 12\um\
luminosity. The 60 and 25\um\ luminosities rise faster than linearly with
bolometric luminosity, while the optical flux rises less than linearly with
bolometric luminosity. This result is a confirmation of the observation that
more luminous disk galaxies have relatively more dust--enshrouded stars.
Increases in the dust content shifts luminosity from the optical to 25---60\um,
while leaving a ``pivot point'' in the mid--IR essentially unchanged. Thus
12\um\ selection is the closest available approximation to selection by a
limiting bolometric--flux, which is approximately 14 times $\nu L_{\nu}$ at
12\um\ for non--Seyfert galaxies. It follows that future deep surveys in the
mid--infrared, at wavelengths of 8 to 12\um, will simultaneously provide
complete samples to different bolometric flux levels of normal and active
galaxies, which will not suffer the strong selection effects present both in
the
optical--UV and far--infrared.

\end{abstract}


\newpage

\section{INTRODUCTION} \label{intro}

In Spinoglio \& Malkan (1989---SM hereafter), we showed that the 12\um\ flux is
approximately a constant fraction of the bolometric flux in active galactic
nuclei (AGNs) of all types (e.g., Quasar/Seyfert~1, Seyfert~2, Blazar). We
therefore used the IRAS sky survey to select an all--sky ($|b|\geq25^\circ$)
sample of galaxies to a limiting 12\um\ flux, thereby providing an unbiased
sample of AGNs. Recently, we used the {\it IRAS Faint Source Survey\/} data
(Moshir et al. 1991) to extend this work to a fainter 12\um\ flux limit, of
0.22~Janskys. Our resulting extended sample includes 893 galaxies, 120 of which
are known to have Seyfert~1 or 2 nuclei (Rush, Malkan, \& Spinoglio 1993---RMS
hereafter). Note that the constant ratio of 12\um\ to bolometric flux was shown
to apply to measurements dominated by the active {\it nucleus\/}. The IRAS
12\um\ selection is highly effective at discovering Seyfert galaxies because
their hot nuclei usually dominate their total 12\um\ fluxes. (This was verified
by comparison with small--beam 10.5\um\ measurements in Edelson, Malkan, \&
Rieke 1987---EMR hereafter). Until the present paper, however, multiwavelength
determinations of the bolometric flux for a significant sample of {\it
normal\/}
(non--Seyfert) galaxies as a whole were not available.

The mid--to--far--infrared emission by which we selected these galaxies arises
from dust grains which absorb short--wavelength energy and thermally
re--radiate
it. In a more intense radiation field, the grains will have higher equilibrium
temperatures, and will tend to emit more at shorter infrared wavelengths. For
many of the 12\um\ selected galaxies, which are to some extent selected to have
``hot" IRAS colors, much of the IRAS emission, especially at 12 and 25\um\ is
ultimately powered by luminous young stars, which are still embedded in the
dust
clouds out of which they recently formed. And of course in the case of AGNs,
there is an additional source of mid--infrared flux powered by the nonstellar
nucleus. Thus the IRAS fluxes, especially at 12 and 25\um\ provide a measure of
what is {\it unusual\/} about these galaxies. To better understand these
galaxies, we also need quantitative information about their more {\it normal\/}
attributes. In particular, it is very helpful to have good measures of the {\it
underlying\/} stellar population which are relatively independent of the
presence of: a) recent bursts of star formation; and b) internal absorption by
dust. It is well known that these requirements are best met by observations in
the near--infrared wavelengths. The light of normal galaxies in the standard
photometric bands J[1.2\um], H[1.6\um], and usually K[2.2\um] is dominated by
red giants in the quiescent (old) stellar population, which is most closely
tied
to the total mass of the galaxy (Recillas--Cruz et al. 1990, 1991). Thus the
intrinsic JHK colors of most galaxies are confined to a relatively narrow range
(Frogel et al. 1978). Since the extinction in the near--infrared is much lower
than that in the visual, only a small fraction of galaxies are dusty enough so
that their near--IR colors are detectably altered by reddening. In some
galaxies
with exceptionally strong starbursts, or an AGN, a measurable ``excess" above
this stellar photospheric emission is detectable at 2.2\um, and more easily at
3.5\um. We detect both effects in the observations presented below.

This paper presents the results of an extended observing campaign in which we
obtained accurate near--infrared aperture photometry for the 12\um\ Galaxies
(\S~\ref{obs}). We then use these to derive colors and total near--infrared
luminosities, for comparison with their IRAS and other properties
(\S~\ref{apcor} and~\ref{analysis}). We compute the bolometric luminosities for
a large sample of normal and Seyfert galaxies (\S~\ref{bollum}) and present and
discuss their normalized spectral energy distributions (\S~\ref{seds}).
Finally,
we give our conclusions (\S~\ref{summary}).


\section{OBSERVATIONS} \label{obs}

We have collected J, H, K, and in some cases L, photometry on 321 galaxies
during eleven photometric observing runs. Most of the data for northern
galaxies
were obtained with the 2.1~m reflector at the San Pedro Martir Observatory
(hereinafter SPM; Baja California, M\'exico), and for the southern
galaxies with the 1.0~m reflector at the European Southern Observatory (ESO),
both equipped with standard InSb photometers. Some of the seyfert galaxies
observed at SPM, had been included in a sample for variability studies of AGN's
by Carrasco \& Cruz--Gonzalez (1995). Additional photometry for a few galaxies
was also collected in May 1981 on the Mount Wilson 60-- and 100--inch
telescopes
and during March--April 1984. Eight to ten standard stars were observed on each
night, with the scatter about the derived photometric zero points, after
correction for airmass, being typically 0.01 to 0.02 magnitudes. The galaxies
were observed  with variable integration times that yielded maximum internal
errors of 0.008 magnitudes. Although, the uncertainties in our galaxy
photometry
are usually dominated by the accuracy with which the signal could be
``peaked--up" on the strip chart, this procedure was repeatable. Hence we feel
confident that the typical errors obtained for the set of standard stars are
representative of the ones expected for the galaxy sample. Standard stellar
magnitudes, on the Caltech and ESO systems, were taken from Elias et al. (1982)
and from Bouchet, Schmider, \& Manfroid (1991), respectively.

In Table~1\footnote{The three tables referenced herein do not appear in print
in
this paper, but in the AAS CD--ROM Series ``Astrophysics on Disc", Volume~V,
1995.}$^{,}$\footnote{During the infrared observing runs for this project, 14
galaxies were measured which were not included in the {\it final\/} 12\um\
Sample (for various reasons; see discussion in RMS). The data for these objects
are given in Table~2, which is constructed in the same manner as Table~1.} we
present the results, in flux densities (mJy), using the absolute calibrations
of
the SPM (Carrasco et al. 1991) and the ESO (Bersanelli, Bouchet, \& Falomo
1991)
photometric systems. The first column gives the galaxy name; the second gives
its equatorial coordinates from the IRAS Faint Source Catalog (B1950.0); the
third through sixth columns give the J, H, K and L flux densities (where
available); the seventh column lists the diameter of the aperture used; the
eighth one the telescope and observing date and the ninth the object class.

In our analysis below, we have included published J, H, K and L photometry from
a large number of papers, all of which are summarized in the recently updated
``Catalog of Infrared Observations" (Gezari et al. 1993). This adds 399
observations of 215 objects, bringing our database to a total sample of 483
galaxies for which we have complete J, H and K photometry. Fortunately, this
includes virtually complete data for the Seyfert galaxies, which were the
primary motivation for constructing the 12\um\ Sample.

Some of the galaxies were also measured with similar apertures by previous
investigators (e.g. Rudy, Levan \& Rodriguez--Espinosa 1982; Balzano \& Weedman
1981). The typical agreement in magnitudes is 0.07 mag rms, and the agreement
on
infrared colors is even better (1$\sigma$=0.04 mag), since several systematic
sources of error cancel out.

\section{APERTURE CORRECTIONS: CALCULATION OF $H_{-0.5}$ AND TOTAL MAGNITUDES}
\label{apcor}

RMS made a special effort to estimate the {\it total\/} far--infrared fluxes
(representing the full spatial extent) for each of the 12\um\ galaxies. This
extra work, not undertaken in most other studies, was required so that valid
flux comparisons could be made between distant and nearby galaxies.
Calculations
of quantities such as luminosity functions require that our measurement of a
given galaxy's flux should not depend on whether or not it happens to be
resolved by the IRAS beams. Our reported IRAS fluxes thus give good measures of
the far--infrared light of the entire galaxy. In this paper we compare these
with ground--based photometry of the total light from the underlying stellar
population, as measured in the near--infrared. Unfortunately there are
virtually
no direct measurements of the total near--IR fluxes of nearby galaxies.
Instrumental limitations and sky noise dictate that the J, H and K band
measurements were made through circular apertures (6---30 arcseconds) much
smaller than the effective IRAS beams. To allow a direct comparison of the
near-- and far--infrared fluxes we applied aperture corrections to transform
the
former into total magnitudes.


The procedure was straightforward for each of the 274 galaxies which has a
value
of $A_{eff}$ listed in the Third Reference Catalog of Galaxies (De~Vaucouleurs
et al. 1991---RC3 hereafter) This ``effective" aperture, which includes
one--half the total flux from the galaxy--is available for ($\sim$53\%) of our
sample galaxies with NIR photometry. We used the RC3 growth curves, (given in
their Table~11), to extrapolate from the largest aperture used to the total
flux. By comparing the growth curves in RC3 with the observed fluxes in
different apertures, we obtained the best result when we adopted the growth
curve  for the NIR fluxes a curve corresponding to 3 morphological types
earlier
than the type listed in RC3. Evidently the near--infrared continuum for a
galaxy
of given morphological type in our sample is slightly more concentrated than
the
standard RC3 aperture growth curves would predict. For galaxies with $A_{eff}$
but no RC3 morphological type, we assumed an early--type spiral, and in fact
the
differences in the growth curves over the range considered are subtle. The {\it
total\/} corrections ranged from 1 to 3 magnitudes. If they would have exceeded
4 magnitudes, we consider the uncertainty too great to use this procedure, and
we thus used the following method.


For about 391 ($\sim$81\%) galaxies of our sample (including all of those for
which the RC3 gives an $A_{eff}$), the RC3 also gives a corrected isophotal
diameter, $D_{0}$, defined at an isophote of 25.0 blue magnitudes per square
arcsec. For these galaxies, we correct the observed H magnitudes to a standard
aperture, $\log(A/D_{0})=-0.5$, as is often done to obtain a characteristic
measure of most of a galaxy's luminosity. These are the galaxy magnitudes used,
for example, in the near--infrared Tully--Fisher method (Aaronson, Huchra, \&
Mould 1979---AHM hereafter). We have fitted a straight line to the average NIR
magnitude growth--curve of AHM for early--type spirals: ${\Delta}m = m_{obs} -
m_{-0.5} = -1.02 - 2.03 \times\log(A/D_{0})$ for Sa's. For a typical value of
$\log(A/D_{0}) = -0.90$ , the (negative) correction to the observed magnitude
was on average 0.81~mag. To correct these to ``total" values, we examined the
correlation between the $-$0.5 magnitudes and the total magnitudes derived for
the 274 $A_{eff}$ galaxies discussed above. We found that the total magnitudes
were on average 0.9 mag. brighter, with an rms scatter of 0.35~mags. We
therefore subtracted 0.9 mag from the $-$0.5 magnitudes. The resulting total
magnitudes are not more accurate than 0.35 mag. (1 $\sigma$).

Several dozen galaxies have multi--aperture photometry which allows us to
intercompare our estimates of total infrared magnitudes\footnote{Each observing
aperture for a given galaxy, when extrapolated with the RC3 or AHM growth
curves, gives an independent estimate of its total magnitude which, in the
absence of errors, should agree with each other.}. The total magnitude
estimated
from the smaller aperture is not on average different from that estimated using
the larger observed aperture, and the random scatter among 30 pairs of
measurements is $\pm$ 0.25 magnitudes (1 $\sigma$). This scatter is partly a
measure of the consistency of our various sets of multiaperture photometry, and
partly a measure of the adequacy of our assumption that the RC3 growth curves
apply to the near--infrared fluxes of these galaxies. About the same scatter
was
found in comparisons of 60 $-$0.5 magnitudes calculated with the AHM D(0)
technique.

For the Seyfert~2 galaxies, we derived total magnitudes by using the same
extrapolation as for normal galaxies, but only when measured fluxes were
available in large enough apertures that the nuclear component of the galaxy
were no longer affecting the shape of the growth curve as compared to a normal
galaxy. We tested this method by comparing growth curves for several Seyfert~2s
(for which data were available spanning a wide range of apertures) to the
growth
curves from RC3 and found that they were fitting well outside the nuclear
region, which is typically 10---20 arcseconds.

We did not follow this procedure for the Seyfert~1 galaxies, since their
near--IR light is much more strongly centrally concentrated than any of the
normal galaxy growth curves. Instead we used the galaxy/nucleus decompositions
presented by Danese et al. (1992) and Kotilainen et al. (1992), and summed up
the two components to measure the near--IR luminosity. For the 13 Seyfert~1
galaxies in common with the study of Danese et al., we added their estimated
``galaxy" and ``nucleus" fluxes to obtain the total flux at 1.2, 1.6, and
2.2\um. In particular, for Mkn~975, Mkn~9, Mkn~704, Mkn~509, Mkn~530 and
MCG+1--57--16 (2237+07), we have taken the nuclear and galactic magnitudes as
given by them, while for Mkn~335, I~Zw~1, Mkn~618, Mkn~79, Mkn~766, NGC~5548
and
Mkn~817, we have taken the nuclear magnitudes and the K galactic magnitude that
they give and assumed their average $(J-K)_G=1.13$ and $(H-K)_G=0.45$ to derive
galactic J and H magnitudes.



We have 16 Seyfert~1 galaxies in common with the study of hard X--ray selected
galaxies of Kotilainen et al. (1992). Two of these are also in the list of
Danese et al. (1992) (NGC~5548 and Mkn~509), and we adopted their estimates.
For
the 14 remaining galaxies we have used the J and K absolute magnitudes of
Kotilainen \& Ward (1994) for both galaxy and nucleus, and converted into flux
densities at earth using their quoted redshifts, $H_0$ and calibration (Wilson
et al. 1972). For the H band fluxes, we adopted the values of non--stellar
fluxes given by Kotilainen et al. (1992) for the nuclear component, while for
the galactic component, we used the colors in an annulus around the nucleus
(from their Table~2) to scale the galactic colors\footnote{Using $(J-H)_G =
(J-K)_G / (J-K)_A \times (J-H)_A$, where subscript G is for ``galaxy" and
subscript A for ``annulus"}.

The computed total fluxes in the J, H and K wavebands are listed in Table~3,
together with the [J--H], [H--K] and [K--L] colors in the smallest aperture
available and the derived near--infrared, far--infrared and bolometric
luminosities. We assumed $H_o = 75~km~s^{-1}Mpc^{-1}$. The K correction
(generally very small) and the corrections of the observed redshift are
described in RMS. The near--infrared luminosity is derived from direct
integration assuming power--law interpolations between 1.2, 1.6, and 2.2 \um.
Similarly, the far--infrared luminosity is derived by trapezoid--rule
integrations of the IRAS fluxes.

In most of these 12\um\ galaxies the tabulated bolometric (0.4---300\um)
luminosity has been computed by combining the total blue magnitude B$_{0(T)}$,
taken from the RC3, the corrected JHK photometry and the IRAS fluxes. To
account
for the mid--infrared and submillimeter contributions, we used the L band
flux\footnote{If the L band flux was not measured, we estimated it from the
correlation of [K--L] with [H--K] color, as discussed in
Appendix~\ref{app:mags}.} and we extrapolated the 60\um\ and 100\um\ IRAS
fluxes
with grey body emission\footnote{We refer to Appendix~\ref{app:submm} for the
details of this computation.}, respectively. For most galaxies the calculated
bolometric luminosity is 98\% or more of the total luminosity we would have
measured from {\it complete\/} multiwavelength coverage (since, the UV flux
drops rapidly beyond the Ca~II HK break, as shown, for example, in Malkan \&
Oke's (1983) ``STDGAL" spectral energy distribution). However for the Seyfert
galaxies, both UV and X--rays can significantly contribute to their bolometric
luminosities. Therefore we have included a crude power--law interpolation
between the B band (4400\AA), the 1330\AA\ band as measured by the IUE SWP
(Edelson et al. 1995) and the soft X--ray band at 1 KeV as measured by the
ROSAT
All Sky Survey (Rush et al. 1996a; details of these data will be presented in
Rush, Malkan, \& Spinoglio 1996b). For most of the Seyfert~2s, UV and X--ray
detections were not available. Fortunately, we have verified that in those
which
the UV and X--ray emission {\it is detected}, it adds only about 3\% percent to
the 0.4---300\um\ luminosity. This justifies our decision to ignore the UV and
X--ray wavebands in the calculation of all Seyfert~2, LINER and starburst
bolometric luminosities.

\section{ANALYSIS: CORRELATIONS OF COLORS} \label{analysis}

\subsection{Near Infrared Colors} \label{analysis_nir}

Figure~1{\it{a}} shows the [J--H] vs. [H--K] two--color diagram for all
available measurements (483) of the 12\um\ Sample of Galaxies. In all of the
diagrams filled squares represent Seyfert~1 galaxies, open squares Seyfert~2
galaxies, asterisks high IR--luminosity non--Seyfert galaxies which we refer as
to ``starburst galaxies" (see the discussion in RMS), open circles ``LINERs"
(Low Ionization Nuclear Emission Region galaxies; Heckman 1980) and diagonal
crosses represent ``normal" (i.e., all other) galaxies. In this and following
diagrams, when a near--IR color is plotted (i.e., and combination of J, H, K or
L magnitudes), the data---referred to simply as ``nuclear"---refer to the
smallest aperture available, since this is usually the one with the best L
magnitude coverage (due to the lower sky noise). The smaller aperture
accentuates the difference between galaxies with and without active nuclei. In
the near--infrared color/color diagrams, vectors show the effects of 1 and 5
magnitudes of visual extinction. This is for an external screen obeying a
standard interstellar reddening law (Rieke \& Lebofsky 1985). In reality the
dust and starlight are probably intermixed, resulting in higher reddening being
inferred at longer wavelengths.


The region of ``normal galaxy colors" is indicated in Figure~1{\it{a}} with a
circle whose radius is equal to 2$\sigma$, centered on the average colors of
$\exvalt{J--H}=0.77\pm0.008$ and $\exvalt{H--K}=0.34\pm0.007$ (uncertainties
represent the standard deviation of the mean; the $1\sigma$ scatter in
individual data points are 0.12 and 0.15, respectively; in fact an ellipse
would
be somewhat more appropriate due to the positive correlation between these two
colors). These values are redder than the average colors of normal galaxies
measured by Frogel et al. (1978) and Recillas--Cruz et al. (1990, 1991), of
$<$J--H$>$ = 0.72 and $<$H--K$>$ = 0.22. Although the effect is subtle and the
scatter is large, the inconsistency is that there is about 12\% excess K flux
in
our sample compared to the others. These ``2.2\um\ excesses" may be
attributable
to thermal emission from extremely hot dust, presumably in actively
star--forming regions (since our galaxies have higher SFRs than other
``normals"
on average; as, for example, Joseph et al. (1984) found K--excesses in
starburst
galaxies), as well as a possible contribution from red supergiants.
Nonetheless,
within our sample, most (90\%) of the non--Seyferts lie within the ``normal"
circled region. A small number of the non--Seyferts are substantially redder in
[H--K] {\it and\/} [J--H], and appear to lie along the reddening vector shown
in
the lower right corner. Thus these galaxies could have intrinsically ``normal"
colors, with the addition of large line--of--sight extinctions (of typically
3---5 magnitudes in $A_{V}$). Only about 10 of the non--Seyferts fall
significantly to the right of the normal--galaxy region, indicating that very
few normal galaxies have strong 2.2\um\ excesses (again, probably due to hot
dust or red supergiants), and about half of these have high infrared
luminosities. One galaxy only, NGC~3353 (=~Mkn~35), has a [J--H] color which is
much bluer than the average (see Figure~1{\it{a}}) . This is one of the bluer
members of the class of compact dwarf galaxies already known to have abnormally
blue colors in the NIR (Thuan 1983).

The Seyfert~nuclei cannot be distinguished from the non--Seyferts in [J--H],
but
they tend to be much redder in [H--K]. As is well known, this is because the
red
giant spectra in normal galaxies have rising flux densities from 1.2 to 1.6\um,
and then falling fluxes from 1.6 to 2.2\um. The Seyfert~nuclear continuum has a
flux density which rises monotonically from 1.2 to 1.6 to 2.2 to 3\um, and
therefore stands out better at longer wavelengths. Most ($\sim$ 73\%) of the
Seyfert~1's and nearly half of both the Seyfert~2s ($\sim$ 41\%) and the
starburst galaxies ($\sim$ 44\%) lie outside the 2$\sigma$ circle defined by
the
normal galaxies and have significantly non--stellar [J--K] colors in small
apertures. In agreement with Cruz--Gonzalez (1984), the Seyfert~1s lie
somewhere
along the track which shows a mixture of normal starlight plus varying
proportions of ``quasar light". The redder Seyfert~1 nuclei in this and other
color diagrams are very similar to the 9 brightest PG quasars (Edelson 1986),
whose average infrared colors ($<$J--H$>$ = 0.85 $\pm$ 0.14 $<$H--K$>$ = 0.98
$\pm$ 0.31 $<$K--L$>$ = 1.59 $\pm$ 0.18; Neugebauer et al. 1987) are indicated
by the cross. These are in turn very similar to the average zero--redshift
quasar colors from Hyland \& Allen (1982) of: $<$J--H$>$ = 0.95 $<$H--K$>$ =
1.15. As expected the published estimates of pure normal galaxy colors and pure
quasar colors are very consistent with the two extremes of the IR color
sequence
in Figure~1{\it{a}}.

The nuclear component we identified in Seyfert~1's---pure ``quasar light" has a
flux at 60\um\ only slightly higher than at 25\um. We believe that the nuclei
most if not all Seyfert~1s and quasars have a peak in their flux density at a
wavelength of 30---50\um. In other words, pure quasar light has a peak energy
output, in $\nu\times f_{\nu}$, at wavelengths of 15---20\um. In the less
luminous Seyfert~1s the observed flux at 60\um\ has a substantial contribution
from the host galaxy. Since  the active nucleus produces very little 100\um\
emission, observations at that wavelength instead measure the host galaxy.

An even more sensitive measure of excess infrared flux over that expected from
red giant photospheres is found in the [H--K] vs [K--L] diagram of
Figure~1{\it{b}}, although fewer measurements are available (123), especially
for non--Seyferts (only 36 normal galaxies). The average colors for normal
galaxies (given with 1$\sigma$ individual scatter) are:
$\exvalt{H--K}=0.34\pm0.12$ (as mentioned above), and
$\exvalt{K--L}=0.75\pm0.64$, the latter having the  much larger scatter. Only
one normal galaxy (NGC~5253) (see Figure~1{\it{b}}), lies close to the region
defined by the PG quasars. The average colors of the 39 Seyfert~1's are
$\exvalt{H--K}=0.84\pm0.33$ and $\exvalt{K--L}=1.31\pm0.49$; for the 24
Seyfert~2's they are $\exvalt{H--K}=0.61\pm0.37$ and
$\exvalt{K--L}=01.05\pm0.59$; for the 8 starburst galaxies they are
$\exvalt{H--K}=0.58\pm0.13$ and $\exvalt{K--L}=1.09\pm0.30$; for the 17 LINERs
they are $\exvalt{H--K}=0.32\pm0.11$ and $\exvalt{K--L}=0.38\pm0.30$.


\subsection{Combined Near Infrared and IRAS Colors} \label{analysis_niriras}

The IRAS and near--infrared wavebands can be combined to separate the different
galaxy classes. The ratio between the 25\um\ flux and the K band flux
(hereafter
referred to as the [25--K] color) plotted against the [60--25] color has
already
been shown to discriminate well between Seyfert~1s and Seyfert~2s (EM, EMR). We
show this in Figure~2{\it{a}} for our sample. The [25--K] color alone is
evidently one of the best discriminators of Seyfert~1 and Seyfert~2 galaxies.
As
noted by EM and EMR, the Seyfert~2 mid--infrared slopes are systematically
steeper by about 0.5 than the Seyfert~1s and quasars. Interestingly, the two
luminous ``Seyfert~1" galaxies with the most unusual colors (steep 2---25\um\
slope but flat 25---60\um\ slope) are both powerful radiogalaxies: 3C~234 and
3C~445. It is also evident in Figure~2{\it{a}} that the starburst galaxies tend
to have steeper 2---25\um\ slopes than normal galaxies.

Similar useful results are obtained with the [25--K] vs. [H--K] two--color
diagram (Figure~2b). Here Seyfert galaxies are well separated from LINERs and
that the high--luminosity non--Seyferts are also well separated from LINERs,
although they overlap with the Seyferts having the smallest [H--K] colors.
Again
the starburst galaxies are redder in H--K than normal galaxies, and show some
overlap with the Seyfert~2s but not the Seyfert~1s.

One of the sharpest separations in all of these diagrams is between the colors
of ``LINERs" and those of the ultraluminous IRAS galaxies (our ``starbursts").
In virtually all cases, the LINERs lie within the range of colors defined by
``normal" galaxies, concentrated to the weakest far--IR fluxes. In contrast
many
of the starbursts have significant excesses in the mid--infrared (in fact, from
2.2 up to 25\um). This is strong prima facia evidence against the notion that
LINERs harbor {\it any\/} unusual current star formation. If a recent burst of
star formation were to explain the characteristic LINER emission line spectra,
it would at have to be very different from those studied previously. The energy
from this activity would at the least have to be hidden at mid--IR wavelengths.
There is no evidence for abnormal emission at {\it any\/} infrared wavelength
in
the LINERs in our sample. The extreme colors of the starbursts could be
understood if the 3.5\um, 25\um, and 60\um\ emission in those objects is
affected by emission of small stochastically--heated dust grains, as found in
the Orion Complex by Bally, Langer, \& Liu (1991) and Wall et al. (1995).

The combination of IRAS and near--infrared colors also defines a clear sequence
from normal galaxies and active ones. The [60--25] color versus the [H--L] is
shown in Figure~3. In the upper left of this diagram, normal galaxies and
LINERs
define a locus of low nuclear activity. A few Seyferts and the high--luminosity
non--Seyferts also lie in this region. However, the Seyfert galaxies (both
types) also extend to the lower right part of the diagram, where no normal
galaxy is located, defining a region of high nuclear activity. Thus the steep
far--IR slopes of normal galaxies and high--luminosity non--Seyferts correspond
to the bluer near--infrared spectra, while the flatter far--IR active galaxies
have redder near--IR slopes. As we will discuss in a future paper (Rush et al.
1996b), this infrared trend to more dominant Seyfert activity is mirrored in
several optical observables which shift systematically as the galaxy is found
closer to the lower right corner of the figure. The least squares line between
all data points also fits the Seyfert galaxies alone with similar regression
coefficients (R= 0.67 for all galaxies, R= 0.64 for all Seyferts). This line,
fitting all the data points, happens to connect the average colors of normal
galaxies (at the upper left corner of the figure) exactly to the average color
of the 9 brightest PG quasars selected by Edelson (1986). The high luminosity
non--Seyferts, as can be seen from their average colors, are significantly off
the best fitting line, having an excess in the 60\um\ emission. It follows that
the infrared energy distributions of these ultraluminous IRAS galaxies, which
do
not harbour a ``classical" Seyfert nucleus, cannot be explained by the
combination of a quasar plus a galactic component which fits the Seyfert
galaxies.

It is not surprising that the only Seyferts classified as ``Type~2" in the
lower
right portion of the diagram (near the realm of ``pure quasar light") are the
most extreme members of this class: Mkn~348 and~463, NGC~1068 and~5506,
FSC~00521--70 and Tololo~109. In fact spectropolarimetry, infrared and/or
X--ray
spectroscopy has indicated that most of these AGNs harbor an obscured Seyfert~1
nucleus. This is consistent with our conclusion from Figure~3 that these are
best described as reddened Seyfert~1 galaxies.

The fact that most Seyfert~1s are located in the lower right part of the
diagram
while most Seyfert~2s are displaced in the upper left corner, where only a few
Seyfert~1s are found (as well as most Messier galaxies classified as Seyfert~1s
or~2s by V\`eron--Cetty \& V\`eron (1991)---see RMS), shows that the energy
distributions of the two Seyfert classes are different: the Seyfert~1s are
steeper in the near--infrared and flatter in the IRAS wavebands, while the
Seyfert~2s behave in the opposite way. The mid--infrared wavelengths, from 3.6
to 12\um, are depressed for the ``Type~2" galaxies and enhanced for the
``Type~1" galaxies. It can very well be that the hot nuclear continuum of
Seyfert~1s is either non--existent in Seyfert~2s (if the latter have physically
different nuclei compared to Seyfert~1s), or it is blocked by circumnuclear
material, as hypothesized in a popular version of the unified model. The
mid--infrared flux of the Seyfert~2s would therefore be due partly to the
galactic component, plus---in some cases such as NGC~1068---a starburst
component which steepens both the [H--L] and [60--25] colors. These results are
in agreement with those of Maiolino et al. (1995), who find that the nuclear
10\um\ luminosity of Broad Line Seyfert galaxies, i.e. ``Type~1" galaxies, is
systematically larger than that of those Seyferts without a BLR, i.e. ``Type~2"
galaxies; and that the host galaxies of Seyfert~2 nuclei tend to have a higher
star formation rate than the host galaxies of Seyfert~1s (this is discussed in
more depth in Appendix~\ref{comparison}).

The effect of adding a strong starburst component is to increase the H--L color
without much altering the 60/25\um\ flux ratio (i.e., corresponding to a
horizontal translation to the right in Figure~3). The classical Seyfert
galaxies
which show this effect most strongly (Mkn~231, Mkn~273, NGC~1068, and NGC~7469)
are already known to contain unusually powerful starbursts. On the other hand,
Arp~220 has too small (blue) an H--L color to be consistent with a Seyfert~1
(``buried quasar") as the principal energy source.

Figure~4{\it{a}} shows an even clearer separation between the Seyfert sequence
and the starbursts. The former tend to have ``hot" 12---60\um\ colors, while
the
latter have steep 12---60\um\ slopes. This is because the starburst enhances
both the 25 and 60\um\ continuum, while the Seyfert~nucleus enhances the 25, 12
and 3.5\um\ emission. Also in this diagram the two objects Arp~220 and Mkn~273
have high 60\um\ emission relative to Seyfert galaxies and quasars.

In principle, a good physical definition of a ``starburst" would be: a region
where current rate of star formation substantially exceeds the rate averaged
over its past history. However, translating this into direct quantitative
observational terms has proven difficult. We have therefore simplistically and
arbitrarily designated all galaxies with infrared luminosities above $1.5
\times
10^{11} L_{\odot}$ as ``starbursts".  We find that these galaxies have colors
which differ systematically from those of the ``normal" galaxies. Nonetheless,
the ``starbursts" appear to be simply the extreme tail of the distribution of
normal galaxies.  The starbursts have the largest relative proportions of young
stars and warm dust, but there is no clear dividing line and no qualitative
distinction.

Figure~4{\it{b}} shows the [H--K] color plotted versus bolometric luminosity.
At
the highest bolometric luminosities, there is a clear separation between the
Seyferts and the high--luminosity non--Seyferts. In this diagram, like the two
previous ones, Arp~220 appears as an extreme starburst galaxy, while the
high-luminosity objects with dominant nuclear activity have much redder [H--K],
like the PG quasars, which have [H--K] = 0.98 (see Figure~4{\it{a}}) .

The basic hypothesis of the ``Unified Scheme" for Seyfert~1 and 2 nuclei is
that
these are intrinsically the same physical objects, except that the former are
seen face--on, while the latter are viewed closer to edge--on. On this
hypothesis, therefore, a sample of Seyfert~1's should not differ from a sample
of Seyfert~2's in any emission which is {\it isotropic\/}, such as
low--frequency radio or [OII]~3727 line emission. Figure~5 shows the [25--K]
versus [60--12] color--color diagram which includes Seyfert galaxies and high
luminosity non--Seyferts. This diagram, as already shown in EMR for the CfA
Seyfert galaxies, segregates  Seyfert~1s in the lower left part of the diagram.
Only 5 out of 52 Seyfert~2s lie in this region, three of which are close to the
edge, while 30 out of 48 Seyfert~1s are inside this region. One version of the
Unified Scheme could explain this, if those objects with an edge--on torus, the
Seyfert~2s, have steeper infrared energy distributions.  This would make both
[60--12] and [25--K] redder, because the high frequency radiation is absorbed
by
the optically thick torus and re--emitted at lower frequencies (Pier \& Krolik
1993). Their best--fit model to the small--aperture photometry for NGC~1068 is
also shown in this diagram, as are the average colors for normal galaxies in
our
sample and the colors of various mixtures of the two (see figure). According to
this scheme, our whole--galaxy colors for NGC~1068 appear to have roughly a
30---40\% contribution from the galaxy at 12\um.

\section{ESTIMATING BOLOMETRIC LUMINOSITIES} \label{bollum}

\subsection{Multiwavelength Properties and Twelve--Micron Selection}
\label{bollum_12um}

As discussed above (and well known since the early days of IRAS), normal
galaxies display an enormous range of properties when viewed over a wide
spectral range (from optical to IRAS wavelengths). Classical
optically--selected
galaxies emit only a modest fraction of their bolometric luminosity in the
thermal infrared (e.g., $\log(L_{FIR}/L_B \sim$--0.3~---~--0.4 for
Shapley--Ames
galaxies---De~Jong et al. 1984). In contrast, IRAS discovered some galaxies
which are very faint optically, and emit most of their large luminosities
through re--radiation by dust grains. Thus a rule similar to that noted by SM
for Seyfert~nuclei may well apply to disk galaxies: they define a sequence of
increasing dust content. As the proportion of dust increases (or intercepts an
increasing fraction of the primary stellar radiation), the galaxy becomes
fainter in the optical and brighter in the infrared. Or even for a fixed dust
content, if a larger proportion of stars (probably younger ones) are embedded
in
warm dust, a growing fraction of the galaxy's bolometric output will be
transferred particularly to the 25\um\ and 60\um\ bands. Since the old stellar
population in the IR--Luminous galaxies is not unusually luminous, these
galaxies are not extremely massive. Instead their high luminosities are
probably
the result of a temporary increase in the current star formation rate which
results in an unusually low mass/light ratio. This could well be a phase that
many normal galaxies pass through, with a relatively low duty cycle. It has
already been noted that this duty cycle could coincide with the fraction of the
lifetime spent by a normal galaxy in close encounters with other galaxies.

As SM found for Seyfert~nuclei, it is also plausible that some intermediate
waveband is relatively unaffected by these processes. In other words, for a
wide
range of normal galaxies of the same bolometric luminosity, there could be a
crossover wavelength in the energy distributions where a relatively constant
proportion of the luminosity is emitted. Our new data suggest that this occurs
and, as is the case for Seyfert nuclei, the crossover wavelength is in the
mid--IR near 12\um, the wavelength at which our sample was defined.

We note that 12\um\ selection is biased in favor of objects which emit strongly
at 12\um, just as every flux--limited sample is biased at the selection
wavelength---{\it by definition\/}. The particular strength of 12\um\ selection
is that it is biased in favor of a homogeneous group of galaxies, for which the
12\um\ flux is linearly proportional to the bolometric flux. In other words,
our
sample will preferentially include objects which are stronger at mid--IR
wavelengths, but these same objects also have higher bolometric luminosities
and
{\it should\/} be preferentially included in any sample complete to a given
bolometric flux. The tight correlation between 12\um\ flux and bolometric flux
for our different classes of galaxies (SM; this
work---\S~\ref{bollum_lumplots})
is an empirical relationship that holds true {\it regardless\/} of the
particular mechanisms (e.g., hot non-thermal nuclear continuum, thermal
emission
from dust heated by starbursts) which elevate the 12\um\ flux. Thus, although
the spectral energy distributions of a wide variety of galaxies show a wide
range of shapes (from the bluest objects which are strong in the
optical/near--IR and weak in the far--IR, to objects which show the opposite
characteristics), they converge best near 12\um (see Figure~1 in SM). Thus, the
mid--IR is the ``pivot" point where the fraction of the bolometric luminosity
emitted is roughly constant.

Within our sample, this is not only the case for Seyfert galaxies, late--type
spirals, and bolometrically luminous galaxies, but also for galaxies in
general,
including early--type spirals and E/S0 galaxies. Even though the average
fraction of the bolometric flux emitted at 12\um\ is lower for E/S0 galaxies as
a general class (one the order of a percent---Knapp et al. 1989; Mazzei \&
De~Zotti 1994; Mazzei, De~Zotti, \& Xu 1994) than for other galaxies, the
unusual E/S0 galaxies {\it in our sample\/} have the same relation between
$L_{12\mu{m}}$ and $L_{bol}$ as the entire sample
(\exvalt{$L_{12\mu{m}}/L_{bol}$}$~\sim7$\%~ and
$L_{12\mu{m}}\propto{L_{bol}^{1.08}}$---see \S~\ref{bollum_lumplots} and
\S~\ref{seds}), so that those objects which {\it do\/} get into our sample
still
have their bolometric flux well represented by the flux at 12\um. The more
subtle difference between the E/S0s and other galaxies in our sample is
exemplified in Figure~6, where we plot the [B--H] vs. [K--12\um] colors for our
sample. The X's represent the average values for all galaxies and for the E/S0
galaxies (with dotted circles representing 1$\sigma$ individual scatter). The
E/S0 galaxies are roughly evenly distributed over most of the range in which we
see spirals, {\it except\/} for the lower right part of the graph. The average
values of $\log(F_{12\mu{m}}/F_{2.2\mu{m}})$ and [B--H] are 0.75 and 3.85 for
our E/S0 galaxies, as compared to 1.12 and 3.53 for our entire sample,
reflecting the different dust content and, to a lesser extent, the stellar
populations of these types of galaxies. Thus, the fact that our E/S0s still
obey
the same relation between 12\um\ and total luminosity as do the other galaxies
in our sample, while being much stronger in the near--IR and optical, simply
points to the fact that the E/S0s represent the extreme end of far--IR--weak
and
optical/near--IR--bright galaxies mentioned above.

\subsection{Individual Versus Bolometric Luminosities} \label{bollum_lumplots}

Figures~7{\it{a}}---{\it{d}} show the correlation of 12, 25, 60 and 100\um\
luminosities with bolometric luminosity for the non--Seyferts. Deviations from
linearity are present for 25\um, 60\um\ and to a lesser extent for 100\um, but
not for 12\um. Since the ``warm excess" luminosity emerges at {\it both\/} 25
and 60\um, the galaxies of higher luminosity do not appear to have unusual
[25--60] colors. This is not an artifact of the redshift since it is present
for
60 and 25 but {\it not\/} 12\um.

For completeness, we also present the same correlations for Seyfert galaxies in
Figures~8{\it{a}}---{\it{d}}. However, we are fully aware that, since the
bolometric luminosities computed here refer to the whole galaxy, the correct
fit
for the Seyfert galaxies will have to await for total {\it nuclear\/}
luminosities, from which starlight had been removed. This will be done in Rush
et al. (1996b) by using optical photometric and spectroscopic data, and it is
beyond the scope of the present paper. As can be seen from Figures~8{\it{a}},
for all Seyferts together the coefficient of the correlation of 12\um\
luminosity versus bolometric luminosity is 1.09 (with regression R = 0.95, for
47 objects). For Seyfert~1s only, this is close to unity (1.03 with R = 0.97
for
19 objects), while for Seyfert~2s it is higher (1.16, with R = 0.93 for 28
objects). At higher wavelength luminosities increase slightly faster with the
bolometric luminosity (see the figures for the slopes and regression
coefficients).

The near infrared luminosity, obtained by integrating the total J, H and K
magnitudes, as described in \S~\ref{apcor}, is also a good measure of the
bolometric luminosity for normal galaxies (see Figure~9{\it{a}}). However it is
clear from the figure that it does not hold for the high luminosity
non--Seyferts, which show a relative lack of near infrared emission. For all
Seyfert galaxies together, the near infrared luminosity increases almost
linearly with bolometric luminosity (see Figure~9{\it{b}}). While Seyfert~1s
have a linear correlation, Seyfert~2s have a significantly flatter slope (0.84
with R = 0.94 for 29 objects), indicating that the low bolometric luminosity
objects have a near infrared excess, due to a greater contribution from
starlight.

The blue luminosity, for normal galaxies, rises slower than linearly with the
bolometric luminosity (see Figure~10{\it{a}}); again the high luminosity
non--Seyfert galaxies deviate the most from linearity. For Seyfert galaxies
(Figure~10{\it{b}}) the blue luminosity---bolometric luminosity relation is
even
flatter, indicating that a strong bias is always present in optical surveys.

\section{NORMALIZED SPECTRAL ENERGY DISTRIBUTIONS} \label{seds}

For most of the galaxies of our sample, we have computed the spectral energy
distributions (SEDs) normalized to the total fluxes in the frequency range
$12.5<\log{\nu}\mbox{[Hz]}<15.0$. Specifically, we have considered all galaxies
in the 12um sample for which were available fluxes (or magnitudes) in the
following wavebands: B, J, H, K, L, 12, 25, 60, 100. We integrated by
connecting
each waveband with a local power--law slope, to derive a total
4000\AA---300\um\
flux which is essentially the bolometric flux. In each non--Seyfert~1 galaxy
the
fluxes in each waveband were then normalized by its total flux. Energy
distributions of the Seyfert~1 galaxies were normalized by including the
1330\AA\ and soft X--ray wavebands. The 1330\AA\ flux is the average of all
archived IUE SWP spectra for each Seyfert~1, from Edelson et al. (1995) and the
X--ray flux is from the Rosat All Sky Survey results in Rush et al. (1996b).
These soft X--ray fluxes were extrapolated out to 30 keV assuming a $-$1 energy
index. For an additional comparison the average normalized SEDs (also including
the UV and X--ray wavebands) of the 6 brightest PG quasars, for which IUE
spectra were available, has also been computed (Edelson 1986).

We have used IUE observations from Edelson et al. (1995) to test our assumption
that the shorter wavelength power can be safely neglected in all the galaxies
aside from the Seyfert~1s. In the 17 Seyfert~2s observed by both IUE and ROSAT,
the UV+X--ray spectrum contributes from 5 to less than 1\% of the total
(bolometric) flux. In two high--luminosity IR starburst galaxies detected by
IUE, their UV flux is 1---2.5\% of their bolometric flux. In the one 12 Micron
Sample LINER detected by IUE---NGC~1052---the UV flux is 5\% of the total.
Neglecting these contributions therefore has added negligible error to our
bolometric flux estimates.

Figure~11 shows the average normalized energy distributions of normal galaxies
(mean of 288) Seyfert~1s (mean of 19), Seyfert~2s (mean of 29),
high--luminosity
non--Seyferts (mean of 19), and LINERs (mean of 15). In these plots the
vertical
(flux) scale is {\it linear\/} so that the integrals correspond to the usual
total power. This is a useful way to visualize which wavebands make the
principal contributions to the bolometric luminosity. For comparison, the
average energy distribution of the 6 PG quasars is also given. The
normalization
for these quasars has been computed from optical through infrared data from the
literature and by assuming a ratio of UV+X--ray to total flux of 0.08, which is
derived from the average of the 5 Seyfert~1s,\footnote{Namely: IZw1, Mkn9,
Mkn704, Mkn1040 and N5548.} for which we have a measure of the bolometric
luminosity, and that were lying close to the PG quasars in the [60--25] versus
[H--L] diagram (Figure~3). In Figure~12 we show the individual SEDs for all
galaxies where we have calculated bolometric luminosities (except for the
normal
galaxies, which are too numerous to plot individually).

In Figure~12{\it{a}} we show the median and first and third quartiles of the
288
normal galaxy spectra. The scatter in the SEDs of normal galaxies is smallest
in
the 8--12\um\ region. Stating this another way, all normal galaxies emit about
7$\pm$1\% of their bolometric flux at 12\um\ (as measured by $\lambda
F\lambda$). This means that a galaxy sample complete to a given 12\um\ flux
limit is also nearly complete to a given bolometric flux limit (of $\sim$14
times the 12\um\ flux limit).

In Figure~12{\it{b}}, we show the normalized SEDs of the 22 Seyfert~1 galaxies:
all objects have similar behavior, with flatter energy distributions than any
other category. In the near-- to mid--IR ($13<\log{\nu}\mbox{[Hz]}<14$) they
are
relatively brighter than any other galaxy type of our sample galaxies. Only the
average spectrum of the bright PG quasar is similar to the Seyfert~1s, and even
brighter at $13<\log{\nu}\mbox{[Hz]}<14$.

In Figure~12{\it{c}}, we show Seyfert~2s, the SEDs of which differ from those
of
Seyfert~1s in two main ways: (1) slightly more 60---100\um\ flux, as is
expected
from dustier objects which reprocess more of the lower--wavelength radiation
into the far--IR; and (2) higher relative fluxes at J and H, i.e. the
Seyfert~1s
have a greater average 2.2---3.5\um\ excess from a strong AGN, as also
indicated
in the color--color diagrams in Figure~1 above.

In Figure~12{\it{d}}, the high--luminosity non--Seyferts are shown, the shape
being similar for all objects. They carry the same fraction of flux at 25\um\
compared to Seyfert~1s, however they drop down by almost one magnitude at
12\um\, having the same fractional flux as the normal galaxies; they show the
weakest optical emission of any class. This is because so much of their
bolometric flux emerges in the far--infrared, and their optical continua are
heavily reddened. These are both caused by large quantities of dust intermixed
with large numbers of luminous stars. Note that their SEDs intersect those of
normal galaxies around 12\um: both types emit about 7\% of their bolometric
luminosity in that waveband. To consider an extreme example of such
high--luminosity galaxies, we compared these objects to the ultra--luminous
IRAS
galaxy F10214+4724, which also has a spectrum that peaks in the infrared. This
object's IRAS fluxes densities (Rowan--Robinson et al. 1991) increase steeply
at
least up to 100\um\ ($\log(F_{\nu100\mu{m}}/F_{\nu60\mu{m}})\sim0.48$),
corresponding to a rest wavelength of $\sim$30\um\ (with $z\sim2.29$---Brown \&
vanden Bout 1991), indicating that the emission is dominated by stellar
processes similar to that from the much closer high--luminosity galaxies in our
sample.

In Figure~12{\it{e}} the LINERs are shown. Again, the SEDs are very similar,
most of the energy coming from the standard population of red giants
characteristic of early--type galaxies. Small to negligible energy is produced
by reradiation from cool dust grains. There is no evidence at any infrared
wavelength of any flux that could be associated with a nonstellar nucleus. The
LINERs have the lowest emission at infrared frequencies and the highest in the
visual. In other words, they have the most ``normal" infrared properties of any
galaxy subgroup. Their spectra also show a wavelength of minimum
scatter--around
6 to 8 \um. The energy they emit at 12\um---$\nu L_{\nu}$---is a constant 5\%
of
their bolometric luminosity for all 12 of our LINERs.

\section{SUMMARY AND CONCLUSIONS} \label{summary}

We have calculated bolometric luminosities and normalized spectral energy
distributions for galaxies in the the 12\um\ Galaxy Sample, finding this sample
to be approximately complete in normal galaxies down to a well--defined {\it
bolometric\/} flux limit of $\nu F^{lim}_\nu/0.07 = 1.1\cdot
10^{-10}$erg~s$^{-1}$~cm$^{-2}$, given our 12\um\ flux limit of 0.22~Jy. Future
deeper surveys conducted at wavelengths of 8 to 12\um\ will {\it
simultaneously\/} provide samples of normal and active galaxies which will be
complete to well--defined bolometric flux limits. The normal galaxy sample will
be about 5 times larger, but will be about 2 times less deep. The depth of the
Seyfert galaxy sample at low luminosities will be limited by how well the
nucleus can be spatially separated from the host galaxy. We will use
small--beam
mid--infrared photometry to make this separation for the 12um Seyferts in an
upcoming paper (Rush et al. 1996b). The future surveys at these mid--infrared
wavelengths will not be biased in favour or against particular types of
galaxies
(e.g. blue quasars, starbursts, etc.) because the mid--infrared emission for
any
class of objects is that most closely proportional to the bolometric luminosity
(whereas the far--infrared/optical luminosities rise faster/slower than
linearly
with bolometric luminosity).

The near-- to far--infrared energy distributions, as shown by the [H--L] and
the
[60--25] colors, define a clear sequence of increasing nuclear activity from
normal galaxies, having blue [H--L] color and cold [60--25] color, to those
Seyfert galaxies which have similar properties of quasars, with red [H--L]
color
and hot [60--25] color. Those Seyfert galaxies having a strong galactic
component---relative to the nuclear component---have intermediate colors.
However, the high luminosity non--Seyferts (we include Arp~220 in this class)
lie significantly off the ``nuclear activity" sequence, indicating that their
relative energy distribution cannot be explained by the mixture of galaxy plus
quasar light, but---most probably---only by the occurrence of violent bursts of
star formation which simultaneously flattens [60--25] and steepens [H--L].
Combining near-- and far--infrared photometry is therefore a powerful
diagnostic
for identifying the emission mechanisms (quasars and/or star formation) which
are ultimately responsible for the emitted light of galaxies.

We also find that Seyfert~1s and Seyfert~2s are separated on infrared
color--color diagrams. This can be reconciled  with  ``Unified Schemes" in
which
Seyfert~1s are seen face--on, Seyfert~2s are seen edge-on, and thus have much
of
their higher frequency radiation absorbed by an intervening torus, resulting in
redder colors.

\acknowledgements

We thank the IRAS (Infrared Astronomical Satellite) Team and the staff at IPAC
(Infrared Processing and Analysis Center, Pasadena, CA), noting that without
the
success of the IRAS mission, this work could not even have been conceived. This
research has made use of the NASA/IPAC Extragalactic Database (NED) which is
operated by the Jet Propulsion Laboratory, California Institute of Technology,
under contract with the National Aeronautics and Space Administration. We also
thank the ESO (European Southern Observatory, Chile) and San Pedro Martir
Observatory (Baja California, Mexico) staff for assisting us during the various
campaigns. We are grateful to K.~Matthews and G.~Neugebauer for support of the
infrared photometer at Mount Wilson, and C.~Beichman and D.~Dickinson for
assistance with the observations. We are also grateful to Marion Schmitz who
gave us the electronic version of the Catalog of Infrared Observations promptly
upon its completion, thus saving us from making a thorough and endless
literature search. We thank Jill Knapp for providing us with electronic copies
of the data tables from Knapp et al. (1989). This work was supported by NASA
grants NAG~5--1358 and NAG~5--1719 and DGAPA/UNAM grant PAPIID:IN103992.

\appendix

\section{ESTIMATED [3.5\um] MAGNITUDES FOR COMPUTATION OF BOLOMETRIC FLUXES}
\label{app:mags}

The dramatic increase of sky noise with wavelength makes it extremely difficult
to obtain large--beam photometry for $\lambda > 3{\mu}m$. Nonetheless, for many
of the brightest 12\um\ Galaxies, and for most of the Seyferts, we measured
fluxes at L [3.5\um] or L' [3.75\um], which were presented in Table~1. For most
non--Seyfert galaxies, however, we lack L photometry. To obtain the bolometric
luminosities in \S~\ref{apcor} above, we {\it assigned\/} these galaxies an L
magnitude based on the correlation between [H--K] and [K--L] shown in Figure~2.
The regression line we show for non-Seyferts is: $$[K-L]=([H-K]-0.13)/0.33$$
This relation shows a scatter of $\pm$0.3 magnitudes, which is only a minor
contributor to the final uncertainties of our bolometric luminosities.


\section{ESTIMATED SUB--MILLIMETER FLUXES FOR COMPUTATION OF BOLOMETRIC FLUXES}
\label{app:submm}

Although our ``$L_{IRAS}$" integrations were stopped at 100\um, it is known
that
substantial thermal flux emerges from cooler dust at longer wavelengths. Since
the flux density often peaks at wavelengths longer than 100\um, the IRAS 60 and
100\um\ fluxes alone, if fitted with a single--temperature component, tend to
under--estimated the total sub--mm flux. Longer wavelength data are needed to
determine the necessary correction. The relatively few measurements in the
far--infrared at $\lambda > 100~{\mu}m$ are sufficient to demonstrate that not
all galaxies show the same spectral shapes beyond 100\um. To compare these
results, we have reduced all of this long--wavelength data to one measure of
the
far--IR spectral turnover: the color temperature of a greybody, for an assumed
dust emissivity $\propto\lambda^{-1}$, which would pass through the 100\um\
flux
measured by IRAS.

As shown in Figure~13, the far--IR/sub--mm color temperature is correlated with
the 60---100\um\ slope. The data are from Rickard \& Harvey (1984), Telesco \&
Harper (1980), Thronson et al. (1990), Stark et al. (1989), Joy et al. (1986),
Hunter et al. (1989), and Smith 1982. Not surprisingly, galaxies which appear
hotter in the IRAS bands also appear hotter in the far--IR/sub--mm, although
the
inferred dust temperatures for the longer wavelengths are systematically lower
than the 60---100\um\ color temperature, because of the increasing contribution
of colder dust at longer wavelengths. We have used the correlation shown in the
figure, $T_{color} = 11.4\cdot (\alpha_{60-100} + 4.67)$K to estimate color
temperatures for each galaxy in our sample. We then added the integrated flux
of
a $\epsilon \propto \lambda^{-1}$ greybody for $\lambda > 100{\mu}m$ to our
IRAS
fluxes in deriving the Bolometric Luminosities of Table~3. The assumed
emissivity law is probably conservative. If emissivity in fact drops like
$\lambda^{-2}$, even {\it less\/} than 20\% of the Bolometric Luminosity
typically emerges at $\lambda > 100{\mu}m $.

\section{COMPARISON WITH OTHER STUDIES} \label{comparison}

After submitting the original version of the present paper, we received
preprints discussing the mid--infrared emission from Seyfert nuclei (Maiolino
et
al. 1995; Giuricin, Mardirossian \& Mezzetti 1995). In this section we show our
results to be consistent with theirs, as we discuss further some differences
between the observed properties of Seyfert~1s and~2s. As mentioned above in
\S~\ref{analysis_niriras}, the main conclusions of Maiolino et al. are, first,
that the nuclear 10\um\ luminosity of Broad Line Seyfert galaxies is
systematically larger than that of those Seyferts without a BLR; and, second,
that the host galaxies of Seyfert~2s nuclei have higher star formation rates
than the hosts of Seyfert~1s. From the average energy distributions, normalized
to the total luminosities, that we computed in \S~\ref{seds} (see Figure~11),
we
find that the average ratio of 12\um\ to total luminosity average is
0.075$\pm$.039 for the normal galaxies, 0.130$\pm$.052 for the Seyfert~1s,
0.097$\pm$.04 for the Seyfert~2s, 0.077$\pm$.027 for the Starbursts (and for
comparison 0.039$\pm$.023 for the LINERs). Although the scatter around the mean
is quite large, the trend is for a decreasing fraction of the {\it total}
luminosity to emerge at 12\um\ going from Seyfert~1s through Seyfert~2s to
normal galaxies and starburst galaxies, and ending with LINERs. If we assume
that the average galactic component of the 12\um\ emission in Seyfert galaxies
is equal for both types to the total emission present in normal galaxies, it
follows that the nuclear 12\um\ component of the Seyfert~1s is brighter than
that of the Seyfert~2s. This is essentially the first conclusion of Maiolino et
al.

We can check their second conclusion by using the normalized 60\um\ luminosity
to infer the star formation rate. For our sample, the average ratio of 60\um\
to
total luminosity is 0.38 for ``starbursts", 0.20 for normal galaxies, 0.19 for
Seyfert~2s, 0.15 for Seyfert~1s, and 0.11 for LINERs. Although the scatter is
again large (the standard deviations of the above ratios range from 0.085 to
0.090) the trend indicates a decrease in the fractional 60\um\ emission from
starburst galaxies, through normal, Seyfert~2s, then Seyfert~1s and LINERs.
This
behavior can also be seen from the [60---12] color for the whole 12\um\ galaxy
sample (i.e., the $\log(F_{60\mu m}/F_{12\mu m})$, as plotted, for example, in
Figure~4{\it{a}}). The more a galaxy is dominated by star formation processes
the redder is its [60---12] color. In fact, the [60---12] color decreases from
starburst galaxies (1.31$\pm$0.21 for 38 galaxies), through normal galaxies
(1.05$\pm$0.22 for 705 objects), LINERs (1.00$\pm$0.25 for 30 objects),
Seyfert~2s (0.98$\pm$0.37 for 65 objects), to Seyfert~1s (0.70$\pm$0.38 for 55
objects). We expect that much of the scatter in this relation will be explained
with further observations, as is already the case for several exceptions to
this
trend among the Seyferts: several Seyfert~2s with the lowest [60---12] values
($<$0.9; NGC~1068, MKN~348, and MKN~463, i.e. the best observed ones) are known
to have an obscured BLR, while three Seyfert~1s with high values of [60---12]
($>$1.2; NGC~1365, NGC~7469, and MKN~231) are already known to have significant
starburst components.

Giuricin et al. (1995) used a compilation of ground-based, small aperture
photometric observations at 10\um\ for 117 Seyfert galaxies. They find that the
10\um\ luminosity distribution of Seyfert~1s extends to greater values than
that
of Seyfert~2s. This is not surprising. In fact, in the 12\um\ galaxy sample
(SM,
RMS), the 12\um\ and 60\um\ luminosity functions of Seyfert~1s extend to higher
luminosities than those of Seyfert~2s. This effect is also observed in Maiolino
et al., who reported a 2$\sigma$ difference between the 10\um\ luminosities of
broad-line and narrow-line Seyfert galaxies.  Giuricin et al. also find that
Seyfert~1s and Seyfert~2s differ significantly in their IRAS 12---25\um\ color,
Seyfert~2s being redder than Seyfert~1s.

We agree with Maiolino et al. that the above results can be reconciled with the
unified models for Seyfert galaxies if the 10\um\ nuclear emission is
moderately
anisotropic. This might imply that the physical size of the 10\um\ source has
to
be no larger than the hypothesized obscuring torus and that this has to be
moderately optically thick in the mid-infrared. We also agree with their
suggestion that some Seyfert host galaxies have enhanced levels of star
formation. In particular, this probably helped some of the Seyfert~2's get into
the 12\um\ sample.

\clearpage


\clearpage

\centerline{\bf FIGURE LEGENDS}

\noindent {\bf Figure 1} --- {\it a}: [J--H] vs. [H--K] color--color diagram.
The normal galaxy colors are indicated by the large circle whose radius is
2$\sigma$. Also indicated is the mean value of the 9 brightest PG quasars (see
text). The extinction of 1 and 5 optical mag. is also given. {\it b}: [H--K]
vs.
[K--L] color--color diagram. In this and all following figures, the filled
squares represent Seyfert~1s, open squares Seyfert~2s, asterisks
high--luminosity non--Seyferts, open circles LINERs and diagonal crosses normal
galaxies.

\noindent {\bf Figure 2} --- {\it a}: [25--K] vs. [60--25] color--color
diagram.
The [25--K] color gives a clear separation of Seyfert~1s from Seyfert~2s (see
the text). {\it b}: [25--K] vs. [H--K] color--color diagram.

\noindent {\bf Figure 3} --- [60--25] vs. [H--L] color--color diagram. The line
represent the least squares fit of all data points. Seyfert~1s and~2s are
labelled with their names. The average colors with 1~$\sigma$ errors are
indicated for normal galaxies, high luminosity non--Seyferts, LINERs and the
brightest PG quasars (see text).

\noindent {\bf Figure 4} --- {\it a}: [60--12] vs. [H--K] color--color diagram.
{\it b}: [H--K] color vs. bolometric luminosity.

\noindent {\bf Figure 5} --- [25--K] vs. [60--12] color--color diagram. The
broken lines at [60--12]=1. and [25--K]=1.35 define a region where almost only
Seyfert~1 galaxies are present. The cross represents the 1 $\sigma$ scatter for
the normal galaxies.  The point in the upper left refers to the best fit model
of an optically thick torus seen edge--on for the nuclear colors of NGC1068
(Pier \& Krolik 1993). The dotted line shows a mixture of those nuclear colors
plus our average galactic colors (the numbers give the fraction of galactic to
nuclear flux at 12\um). For comparison we also show our large aperture data of
the same galaxy.

\noindent {\bf Figure 6} --- [B--H] vs. [K--12\um] color--color diagram.
Elipses
representing the 1--$\sigma$ scatter in each color are centered on the average
value for the E/S0 galaxies and for all galaxies.

\noindent {\bf Figure 7} --- Monochromatic luminosities vs. bolometric
luminosity for normal galaxies and high luminosity non--Seyferts. Also shown
are
the points representing LINERs. The lines represent the least squares fit to
all
but the LINERs points. {\it a,b,c,d}: 12, 25, 60, and 100\um\ luminosity vs.
bolometric luminosity.

\noindent {\bf Figure 8} --- Monochromatic luminosities vs. bolometric
luminosity for {\it Seyferts}. The lines represent the least squares fit to all
data points, except the one relative to Arp~220 (which isn't likely a
Seyfert---see text), whose position is labeled. {\it a,b,c,d}: 12, 25, 60, and
100\um\ luminosity vs. bolometric luminosity.

\noindent {\bf Figure 9} --- Near--infrared luminosity vs. bolometric
luminosity
for {\it{a}}) normals and high--luminosity non--Seyferts, and {\it{b}})
Seyferts. The solid lines represent the least squares fit to all data points.
The dashed and dotted lines in {\it{b}} are the fits to the Seyfert~1s and~2s,
respectively.

\noindent {\bf Figure 10} --- Total blue luminosity vs. bolometric luminosity
for {\it{a}}) normals and high--luminosity non--Seyferts and LINERs, and
{\it{b}}) Seyferts. The lines represent the least squares fit to all data
points.

\noindent {\bf Figure 11} --- Average spectral energy distribution normalized
to
bolometric flux of all classes of galaxies. Also included is the average PG
quasar spectrum (see text).

\noindent {\bf Figure 12} --- Spectral energy distribution, normalized to
bolometric flux, of individual galaxies in the 12\um\ Sample. {\it a,b,c,d,e}:
normal galaxies, Seyfert~1s, Seyfert~2s, high--luminosity non--Seyferts, and
LINERs, respectively. For the normal galaxies, we plot only the logarithmic
spectral energy distribution of the median and first and third quartiles points
at each observed frequency.

\noindent {\bf Figure 13} --- Least squares fit of the spectral index
$\alpha_{60-100\mu{m}}$  as a function of the color temperature, assuming grey
body emission with dust emissivity law $\epsilon \propto \lambda^{-1}$.

\end{document}